# Highly Tunable Two-Qubit Interactions in Si/SiGe Quantum Dots by Interchanging the Roles of Qubit-Defining Gates

Jaemin Park[1], Hyeongyu Jang[1], Hanseo Sohn[1], Younguk Song[1], Lucas E. A. Stehouwer[2], Davide Degli Esposti[2], Giordano Scappucci[2], and Dohun Kim[1]*

*[1]NextQuantum, Department of Physics and Astronomy, and Institute of Applied Physics, Seoul National University, Seoul 08826, Korea*

*[2] QuTech and Kavli Institute of Nanoscience, Delft University of Technology, PO Box 5046, 2600 GA Delft, The Netherlands*

**Corresponding author: dohunkim@snu.ac.kr*

## Abstract

Silicon quantum dot spin qubits have become a promising platform for scalable quantum computing because of their small size and compatibility with industrial semiconductor manufacturing processes. Although Si/SiGe heterostructures are commonly used to host spin qubits due to their high mobility and low percolation density, the SiGe spacer creates a gap between the qubits and control electrodes, which limits the ability to tune exchange coupling. As a result, residual coupling leads to unwanted single-qubit phase shifts, making multi-qubit control more difficult. In this work, we explore swapping the roles of overlapping nanogates to overcome this issue. By reconfiguring the gate voltages, we demonstrate in situ role switching while maintaining multi-qubit control. Additionally, this method improves the tunability of exchange coupling by up to 3.6 times. This strategy reduces unintended single-qubit phase shifts and minimizes the complexity of multi-qubit control, supporting scalable growth with minimal experimental overhead.

Spin qubits based on semiconductor quantum dots[1] have been extensively studied as a feasible technology for implementation in scalable quantum processors, largely motivated by their fabrication compatibility with industrial complementary metal-oxide-semiconductor (CMOS) technology.[2] Several studies have demonstrated the operation of one- and two-dimensional spin qubit arrays,[3–7] high-fidelity single- and two-qubit gates,[8–10] long-range coupling via a superconducting resonator,[11,12] and operation at elevated temperatures, which significantly boosts cooling power and offers advantages for quantum-classical chip integration.[13–18] Building on this progress, high-fidelity initialization, readout, and manipulation of spin qubits have now been demonstrated within a single device, as validated by observing the violation of Bell's inequality.[19] In parallel, the industrially mature fabrication technology has also motivated extensive research toward the high-throughput production of high-uniformity spin qubit devices.[20–22]

The formation of quantum dots is a prerequisite for realizing spin qubits. This involves confining electrons or holes to specific regions by leveraging the properties of the material in combination with the electric potential generated by applying voltage to nano-fabricated metal gates.[23] In particular, silicon quantum dot qubits have been extensively studied, due to the high natural abundance of zero-nuclear-spin isotopes. Furthermore, residual nuclear spins can be further removed via isotopic enrichment to provide a magnetically quiet environment for the spin qubits.[24] With this platform, the overlapping metal gate architecture has been widely adopted to realize tight electron confinement.[4,9,25] In early device implementations, approaching the single-electron regime was a crucial objective, which prompted the fabrication of plunger gates with an appropriate width to ensure their energy controllability and sufficient coupling between quantum dots. In many studies, a tuning strategy was employed in which quantum dots are formed beneath the plunger gates located below the barrier gate layer, a configuration referred to as *conventional tuning* in this work. In this tuning scheme, gates with stronger capacitive coupling to the quantum dots are typically assigned to control the dot energy levels. Studies using this configuration have demonstrated that static tunnel coupling between quantum dots can be well controlled. However, as research efforts have increasingly shifted toward high-fidelity multi-qubit control, which often requires dynamic control of the coupling via barrier pulses,[26] the achievable on/off ratio of the coupling with barrier pulses exceeding 100 and stable operation under large barrier voltages have been reported to be limited in some studies.[4,27] These constraints have motivated alternative approaches, such as exploiting multiple barrier pulses or detuning. Meanwhile, spin shuttling, a potential pathway toward long-range coupling, has also been extensively studied in devices where plunger and barrier nanogates of nearly equal width are used to form and move quantum dots continuously.[28]

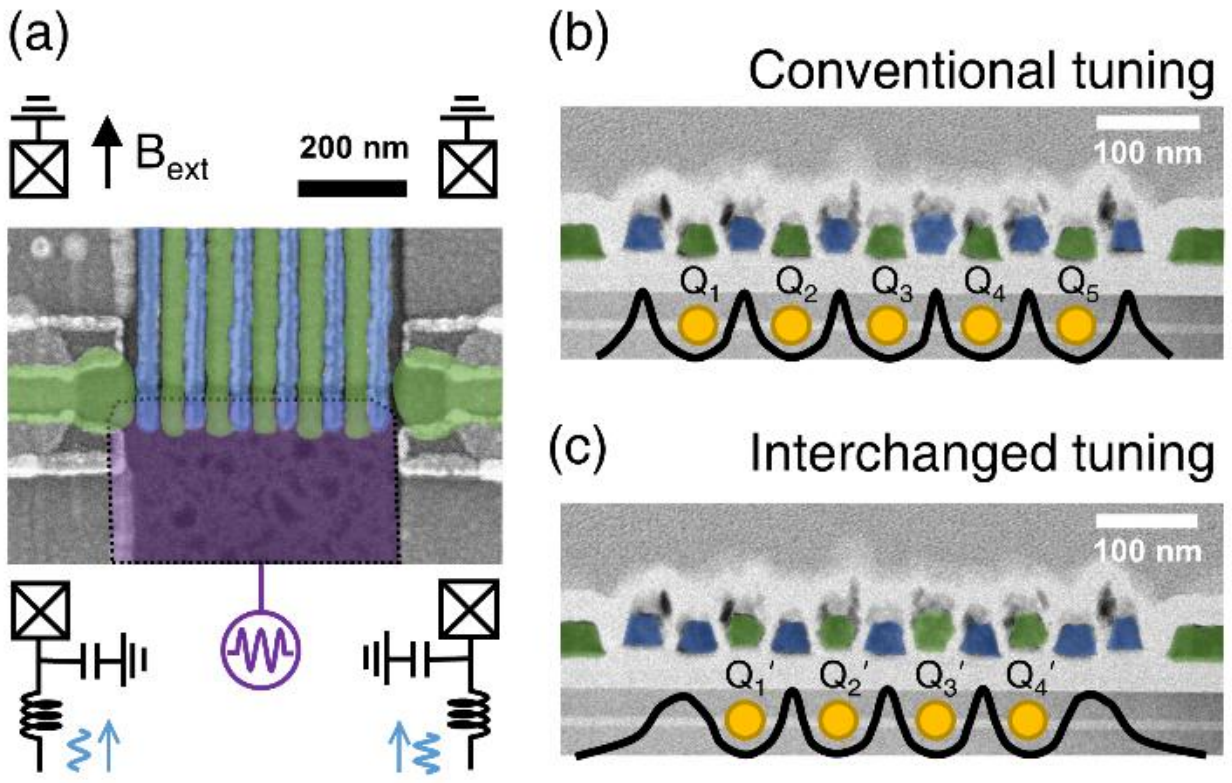

**Figure 1.**

In line with these efforts, we fabricated a quantum-dot qubit device with equal-width nanogates and investigated the effect of changing the tuning strategy. This approach, referred to as *interchanged tuning* hereafter, assigns the nanogates closer to (farther from) the spin qubits as barrier (plunger) gates. Using this tuning method, the well-defined formation of quantum dot single-spin qubits is demonstrated. Although previous studies on shuttling have shown that potential wells can be dynamically formed beneath each gate and moved continuously,[28] shifting multiple wells and energy barriers with maintained multi-qubit controllability has not been demonstrated in a similar gate structure. Here, we show that both of these are feasible with the proposed tuning strategy. Furthermore, the tunability of the exchange coupling is enhanced by several orders of magnitude for all nearest-neighbor pairs with a maximum value of 16.6 dec/V, while the controllability of the spin qubits is simultaneously maintained. We further discuss how this enhancement may influence future spin qubit experiments.

The device was fabricated on an isotopically enriched $^{28}$Si/SiGe heterostructure wafer.[24] First, a 20-nm-thick $Al_2O_3$ layer was deposited by atomic layer deposition (ALD) to serve as insulation, followed by three layers of Ti/Pd metallic gates and oxide layers using E-beam lithography, E-beam evaporation, and ALD. The first of these layers consists of screening gates, which prevent the electric field from accessing the upper metal layers and locally deplete the two-dimensional electron gas (2DEG), thereby roughly defining a one-dimensional channel for quantum dot formation. The second and third layers are used to control either the energy of the quantum dots or the tunnel coupling between them. A micromagnet was fabricated on top of the metal layers to introduce a magnetic field gradient to enable individual qubits to be addressed and to enhance the controllability of spin qubits via electric dipole spin resonance (EDSR). The micromagnet was magnetized by applying an external in-plane magnetic field of 0.44 T in the direction as indicated in Figure 1a. The spin-qubit resonance frequency is typically in the range of 15−16 GHz. Single-qubit operation is achieved by applying microwave signals to the screening gate at the qubit frequency, which causes the spin qubit to experience an oscillating transverse magnetic field.[29] High-bandwidth spin qubit measurements were conducted by employing RF-reflectometry to apply a carrier signal to the ohmic contacts, which are connected to LC tank circuits composed of a commercial inductor and parasitic capacitance.[30] All gates were connected to DC voltage sources and to either an arbitrary waveform generator or a signal generator through on-board bias tees. The device was cooled in a dilution refrigerator, and the electron temperature was approximately 75 mK. Additional details of the experimental setup are provided in the Supporting Information (S1).

Figure 1b shows the gate configuration for the conventional tuning method, along with a schematic of the corresponding potential landscape, where the green (blue) gates located closer to (farther from) the 2DEG act as plunger (barrier) gates. The combination provides a potential energy landscape for an experiment using five quantum dots ($Q_1$, $Q_2$, $Q_3$, $Q_4$, $Q_5$) and two sensing dots for charge sensing, one on each side. Alternatively, the roles of the nanogates can be switched by assigning the role of barrier gates to the gates closer to the 2DEG. The corresponding schematic potential landscape is shown in Figure 1c. This approach allows the formation of four quantum dots ($Q_1'$, $Q_2'$, $Q_3'$, $Q_4'$) with the given gate structure.

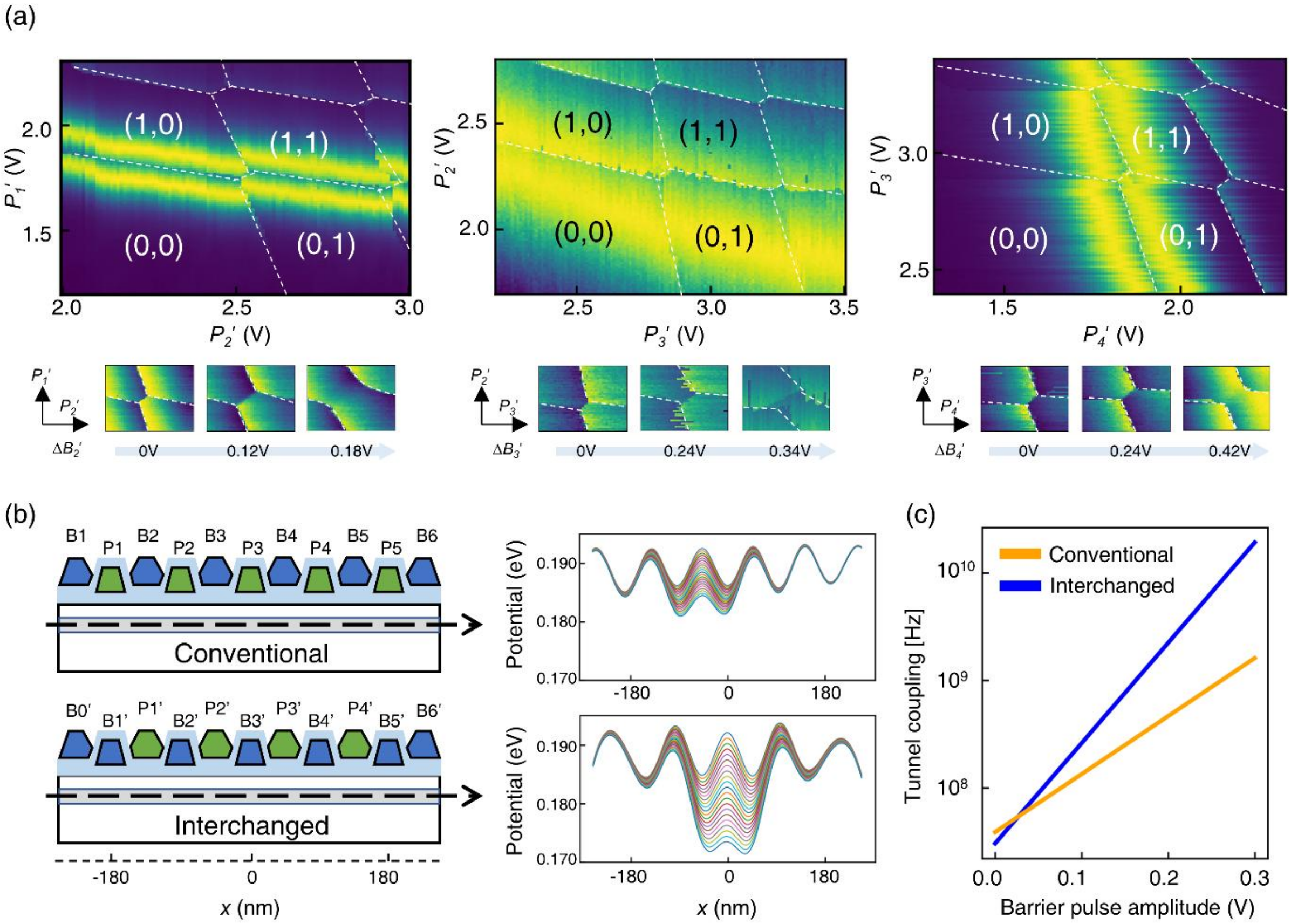

**Figure 2.**

First, the interchanged tuning method is shown to enable the single-electron regime to be reached for all quantum dots suitable for multi-qubit operation. Figure 2a shows the charge stability diagram for the few-electron regime of $Q_1'$-$Q_2'$, $Q_2'$-$Q_3'$, and $Q_3'$-$Q_4'$ with physical gates. Although the lever arm of the newly mapped plungers is reduced, more than three-electron filling is achieved owing to a high-quality 8-nm-thick intergate oxide layer, which prevents breakdown even under a few volts of potential difference between the gates. Parity readout was adopted using the Pauli spin blockade (PSB)[31,32] for spin-state readout, a recent and widely used approach due to its deterministic readout time, applicability to isolated quantum dot qubits, and robust operation at elevated temperatures.[13,33,34] By setting the charge occupation to (3,1,1,3), a wide PSB readout window was realized by exploiting orbital energy spacing rather than valley splitting.[4] The expression $(k,l,m,n)$ denotes the number of electrons in each quantum dot. Under this configuration, the newly designated barrier gates were observed to have the potential to vary the tunnel coupling between dots over a wide range, as qualitatively illustrated in the bottom insets to Figure 2a. Compared with the conventional tuning method, the outermost quantum dots are farther from the sensor dots. Therefore, we used an integration time of 2 μs, which is slightly longer than that of the conventional strategy, to maintain high-fidelity readout. Interchanged tuning produced a signal-to-noise ratio up to 10.6, which corresponds to a charge readout fidelity above 99.9%.[35] (see Supporting Information S2)

The potential profile on 2DEG for the tuning strategies was simulated using COMSOL MULTIPHYSICS (Figure 2b). For both tuning methods, we calculated the potential profiles by varying the voltage applied to one of the additional barrier gates to a maximum of 0.3 V, considering the typical voltage range used in the experiment. As the barrier gate voltage of $B_3$($B_3'$) increases, the potential barrier between $Q_2(Q_2')$ and $Q_3(Q_3')$ gradually decreases, which leads to stronger tunnel coupling. The change is more pronounced in the interchanged gate configuration than in the conventional one. Based on the simulation result, the tunnel coupling was calculated as a function of the barrier gate voltage by solving the single-particle Hamiltonian between the quantum dots. As is evident from Figure 2c, the change in tunnel coupling exhibits a steeper gradient in the case of the interchanged tuning strategy, suggesting that a wider dynamic range of the coupling strengths is accessible.

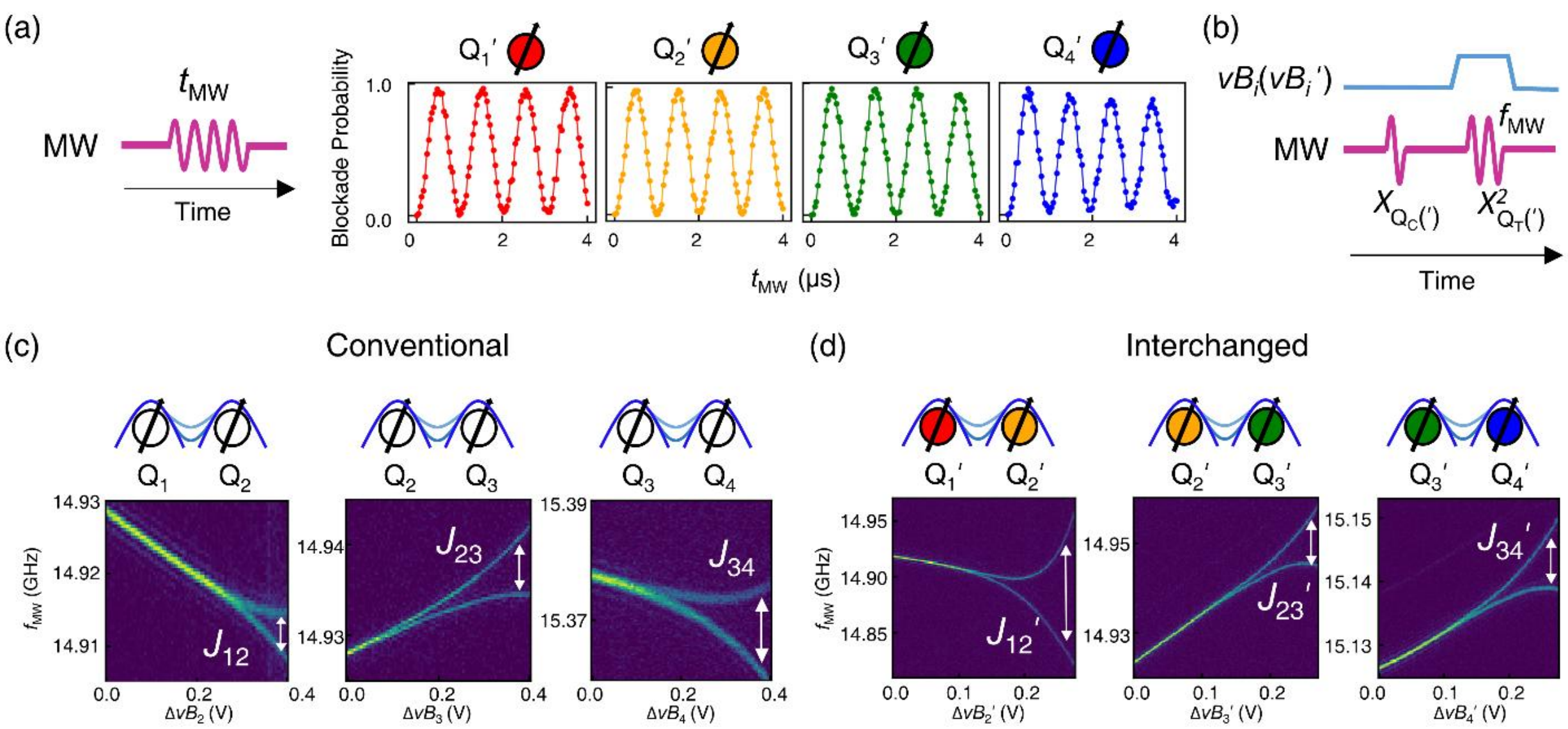

**Figure 3.**

Subsequently, we attempted to experimentally determine how the enhanced tunability of tunnel coupling, as revealed by simulation, translates into the tunability of exchange coupling. The spin qubits were manipulated at the symmetric point in the charge stability diagram, where the first derivative of the exchange coupling ($J$) with respect to detuning ($\Delta$) vanishes (i.e., $\partial J/\partial\Delta = 0$).[26] In this and all subsequent experiments, we employed virtual gates to achieve independent control of each quantum dot's energy, allowing stable operation at the symmetric point. Figure 3a shows that high-quality single-qubit control and addressing can be demonstrated under the interchanged tuning strategy. During the experiments, two pairs of qubits were initialized to odd-parity by adiabatically changing the detuning to move the single electron from the doubly-occupied singlet state to the other quantum dot[4] (see Supporting Information S2). Variation of the burst duration ($t_{MW}$) of the microwave signal applied to the screening gate enabled the high-quality oscillations in the blockade probability to be observed. Although the Rabi oscillation quality of $Q_4'$ was limited, this was attributed to strong coupling with a nearby reservoir.[36]

Exchange spectroscopy was used to quantify the achievable exchange coupling strength. The pulse sequence is illustrated in Figure 3b. First, the control qubit was rotated by π/2 about the x-axis. Then, a virtual barrier pulse $vB_i$ ($vB_i'$) was applied to the gates between the qubits to increase their exchange coupling, and the microwave frequency ($f_{MW}$) was swept to determine the dependence of the resonance frequency on the exchange coupling. Figure 3b and 3c show the trend in exchange coupling, defined as the frequency difference between the two branches in a given exchange spectroscopy map, for the conventional and interchanged gate configurations, respectively. The global slope of the branches with respect to the barrier pulse amplitude arises from the change in location of the quantum dot under the spatial gradient of $B_z$ generated by the micromagnet. For conventional tuning, the exchange coupling increased to only about 10 MHz with a 0.4 V barrier pulse. In contrast, the interchanged strategy could significantly enhance the coupling; for example, $J_{12}'$ reached about 125 MHz at a 0.28 V barrier pulse for $Q_1'$-$Q_2'$. Even at the maximum achievable exchange coupling of 125 MHz, we did not observe unintended transition lines indicative of excessive crosstalk. Although the degree of tunability varied among different qubit pairs, the tunability was clearly enhanced overall.

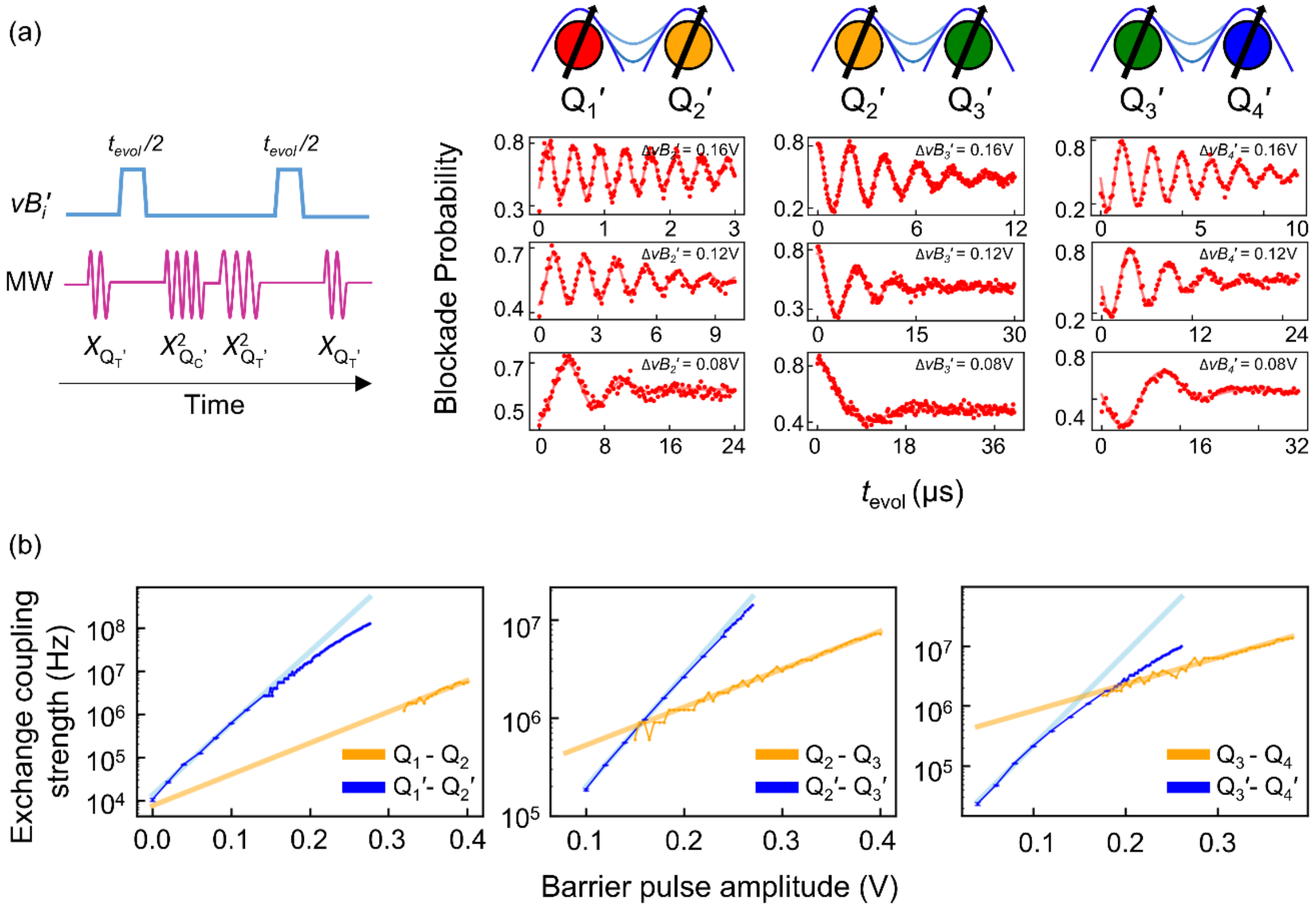

**Figure 4.**

The tunability of the exchange coupling strength was quantified more precisely by measuring the decoupled controlled-Z (dCZ) oscillation.[37] Evaluation of the tunability makes it necessary to examine a coupling strength smaller than the resonance linewidth. Considering that the initial condition without the barrier pulse may already have a relatively large $J$, the time-domain dCZ oscillations were used to verify the coupling behavior. In the pulse sequence in Figure 4a (left), intervening π-pulses ($X^2$) applied to both qubits cancel out the unwanted single-qubit phase accumulation, thus allowing us to observe the phase oscillations by $J$.[37] The phase oscillation data shown in Figure 4a as a function of $t_{evol}$ and $\Delta vB_3'$ can be fitted to a Gaussian decaying sinusoidal function to extract the frequency, which corresponds to $J/2$. The decay time of the evolution ($T_2^{dCZ}$) and the corresponding quality factor can also be extracted. The quality factor, defined as $J \times T_2^{dCZ}$, was evaluated for each qubit pair at various barrier pulse amplitudes. The corresponding values are provided in Table 1. Although the decay time decreases with increasing barrier pulse amplitude, the accompanying increase in coupling strength results in an improved quality factor. A comparison with the conventional tuning strategy over the same range of $J$ shows that the interchanged tuning strategy yields comparable quality factors, and even higher values at some coupling strengths. (see Supporting Information S5). This indicates that the interchanged tuning strategy does not introduce additional charge noise that would degrade the quality factor.

| $\Delta vB_i'$ | $Q_1'$-$Q_2'$ | $Q_2'$-$Q_3'$ | $Q_3'$-$Q_4'$ |
|---|---|---|---|
| 0.08 V | 8.69 | 3.01 | 4.27 |
| 0.12 V | 16.1 | 6.38 | 9.04 |
| 0.16 V | 22.3 | 9.47 | 12.6 |

**Table 1.** Quality factors under the interchanged tuning strategy as a function of barrier pulse amplitude.

Figure 4b compares the exchange coupling tunability between the two tuning methods based on the measured results. With the interchanged tuning strategy, the dependence of the coupling strength on the barrier pulse amplitude is noticeably steeper. The data were fit to an exponential form by focusing on the few-MHz range typically used in experiments. The fitted tunability values are provided in Table 2. Using the interchanged tuning method, the tunability increases by several orders of magnitude for all-nearest-neighbor pairs, significantly exceeding that in the conventional case. Interestingly, the trend of exchange coupling does not follow a single-exponential form over the entire data range. A previous study showed that tunnel coupling asymptotically approaches a constant as the two quantum dots merge into a single larger quantum dot because the potential barrier is lowered by the barrier pulse.[26] This result is itself a consequence of the enhanced tunability of the coupling. This high tunability allows us to explore the entire coupling regime and, henceforth, is expected to help accurately calibrate the actual profile of exchange coupling as a function of the barrier-pulse amplitude. This tuning strategy still allows us to construct virtual gates and suppress crosstalk, thereby enabling reliable multi-qubit control (see Supporting Information S4 and S5). These results indicate that the enhanced tunability can be maintained without introducing crosstalk that is sufficiently large to hinder qubit operation.

| | Conventional | Interchanged |
|---|---|---|
| $Q_1(')$-$Q_2(')$ | 7.25 dec/V | 16.6 dec/V |
| $Q_2(')$-$Q_3(')$ | 3.87 dec/V | 11.4 dec/V |
| $Q_3(')$-$Q_4(')$ | 4.32 dec/V | 15.4 dec/V |

**Table 2.** Fitted tunabilities for the qubit pairs under both tuning strategies.

The most plausible explanation for the increased tunability is a difference in the lever arm. Indeed, when comparing the lever arm ratio from the charge stability diagram (see Supporting Information S2) for each tuning strategy, the ratios are 1.75, 3.18, and 2.71, whereas the corresponding ratios for coupling controllability are 1.71, 2.72, and 2.10, respectively, for the case $J$ ranges from 1 to 10 MHz. These results show that tunability is strongly affected by the lever arm and are consistent with the simulation results presented in Supporting Information S5. The simulation further shows that the lever arm is significantly affected by the bottom width of the gates, which can differ from the intended value due to fabrication-induced modifications. (see Supporting Information S5).

Residual exchange coupling induces unwanted $ZZ$ operation, which complicates the resulting Hamiltonian. When a microwave signal is used to manipulate one of the qubits in a two-qubit system with residual coupling, the resulting time-evolution operator includes not only the single-qubit operation term but also contains an additional term proportional to $\frac{J}{\Omega}ZZ$.[38] Here, $\Omega$ denotes the Rabi frequency. This residual coupling limits the achievable single-qubit fidelity. Mitigation of this effect requires the coupling to be much smaller than the Rabi frequency ($J \ll \Omega$) as a fundamental requirement to achieve high single-qubit gate fidelity. However, to realize both single- and two-qubit operations with high fidelity, the coupling must be dynamically controlled on a short timescale, which is typically challenging to achieve with insufficient tunability. Even though a calibration routine could compensate for the undesired phase accumulation,[8] this becomes increasingly challenging as the scale of the qubit system grows. Therefore, achieving exchange interaction such that it has sufficient tunability is crucial, and requires not only careful design of the gate geometry but also an optimal gate configuration for the experiment. From this perspective, the suggested tuning strategy provides a practical and efficient means to enhance the tunability of the exchange coupling simply by changing the voltage configuration within the same qubit device, without requiring any modification to the experimental setup.

In conclusion, we investigated the effectiveness of an interchanged quantum dot tuning method, in which the roles of the nanogates defining the quantum dots are interchanged. Using the proposed strategy, we demonstrated the formation of quantum dots and single-qubit operation. Notably, the tunability of exchange coupling is enhanced by several orders of magnitude across all nearest-neighbor qubit pairs. The result is significant because it demonstrates the realization of tunability of approximately 16 dec/V in Si/SiGe devices, which has hitherto only been achieved in Si-MOS devices with a larger lever arm.[39] Conventionally, the spin qubits in Si/SiGe devices are located far from the electrically noisy semiconductor-dielectric interface;[24] however, the inherent distance between the qubits and metal electrodes limits the tunability of the coupling. By contrast, the proposed interchanged tuning strategy allows us to effectively exploit the advantages of the Si/SiGe platform, particularly for devices with limited interaction tunability, without degrading the quality of exchange coupling. The enhanced tunability originates mainly from the increased lever arm, consistent with our simulation results. Beyond this relationship, we find that the lever arm strongly depends on the effective bottom width of the gates, and that fabrication-induced changes to the gate geometry can significantly alter the relative influence of the two gate layers on the quantum dots from that intended in the original design. This suggests that not only the gate layout in the top view, but also factors such as the intervening oxide thickness and gate pitch, which modify the gate geometry, should be carefully considered.

In addition, these results show that quantum dot formation is not restricted to a single tuning strategy within a given device architecture, allowing the tuning method to be selected according to experimental requirements. This flexibility allows the device properties to be studied more intensively by varying the location of the quantum dot used as a probe; for example, the magnetic gradient field to optimize micromagnet design, valley splitting as a function of position, and the spatial characteristics of noise originating from impurities and sensor dots. The prior works[40,41] have demonstrated parameter probing using conveyor-mode shuttling. In contrast, our strategy offers an alternative based on a static quantum-dot configuration, compatible with long-duration protocols such as Carr-Purcell-Meiboom-Gill (CPMG) sequences. Although our strategy employs a multilayer gate structure in this work, its underlying principle arises from differences in capacitive coupling between the gates and the quantum dots. Such differences can also be present in advanced single-layer architectures,[21,22] suggesting that the strategy demonstrated here is not limited to multilayer gate structures. In addition, Si-MOS-based qubit devices realized using advanced fabrication techniques often employ overlapping gate structures.[42] This implies that our study is broadly applicable to quantum dot spin qubits across different material platforms, as key quantities related to gate controllability, such as lever arm and exchange coupling tunability, are ultimately governed by the engineering of capacitive coupling between gates and quantum dots, regardless of the host material.

Further studies on fabrication strategies, including thinner gate oxides and advanced gate geometries, would contribute to bolstering the advantages of this strategy. This approach may provide a way to achieve sufficient coupling tunability with modest experimental effort, thereby facilitating convenient multi-qubit control, for example, by optimizing the PSB readout,[43] increasing the efficiency of the resonant SWAP gate operation,[44] and accurately engineering the Hamiltonian for Si/SiGe quantum dot spin qubit processors defined by the gate overlay.

**Supporting Information:** Experimental setup, qubit initialization and measurement methods, simulation of electric potential and tunnel coupling, virtual gate configuration, comparison of exchange coupling quality, and simulation of metal gate geometry on lever arm and exchange coupling tunability. (PDF).

The Supporting Information is available free of charge on the ACS Publications website at DOI: XXXX

## Acknowledgments

This work was supported by a National Research Foundation of Korea (NRF) grant funded by the Korean Government (Ministry of Science and ICT (MSIT)) (RS-2023-00283291, RS-2024-00413957, SRC Center for Quantum Coherence in Condensed Matter RS-2023-00207732, RS-2023-NR077112 and Quantum Technology R&D Leading Program (Quantum Computing) RS-2024-00442994) and a core center program grant funded by the Ministry of Education (No. 2021R1A6C101B418). Lucas E. A. Stehouwer and Davide Degli Esposti developed and characterized the heterostructure under the supervision of Giordano Scappucci.
Correspondence and requests for materials should be addressed to DK (dohunkim@snu.ac.kr).

## Figure captions

**Figure 1.** $^{28}$Si/SiGe linear quantum dot qubit device and schematic representation of tuning strategies

**(a)** Scanning electron microscopy image of the device. The squares with X symbols denote an ohmic contact through which a radio-frequency (RF) signal is injected via a tank circuit for RF reflectometry. The purple gate enclosed within the dashed line represents the screening gate on the first metal layer, to which microwaves are applied for single-qubit manipulation. The gates colored green (blue) correspond to the second (third) metal layer. The micromagnet fabricated on top of the gates is omitted for clarity.

**(b-c**) Transmission electron microscopy image showing a cross-section of a similar device fabricated in the same batch. The estimated electrostatic potential landscape and quantum dot locations for both tuning strategies are shown. For each strategy, the plunger and barrier gates are colored green and blue, respectively.

**Figure 2.** Charge stability diagram of double quantum dot pairs and coupling

**(a)** Charge stability diagram of pairs of double-quantum dots. $P_i'$ denotes the voltage on the plunger gates $\mathrm{P_i}'$ in the interchanged configuration. Insets: variation of charge stability diagram near interdot tunneling as the voltage applied to the barrier gates becomes increasingly positive, with the text in each inset indicating the voltage increments. The label $\Delta B_i'$ denotes the magnitude of the voltage increase on barrier gates. For $Q_2'$-$Q_3'$, the tunneling rate to the reservoir is slow compared with the time scale of voltage rastering, so that the transition lines appear latched.

**(b)** Simulated electric potential for the conventional and interchanged gate configurations. Left: device cross-section for the two tuning strategies. Right: corresponding simulated potential profile at the qubit location as the additional voltage on B3 (B3′) is varied from 0 V to 0.3 V. The data for the conventional tuning strategy are vertically offset for clarity of comparison.

**(c)** Tunnel coupling calculated from the simulated data for both tuning strategies.

**Figure 3.** Single-qubit controllability and exchange coupling spectroscopy

**(a)** Rabi oscillations of four qubits and pulse sequence in the interchanged tuning method as a function of the microwave duration $t_{MW}$.

**(b)** Pulse sequence for exchange spectroscopy.

**(c-d)** Exchange spectroscopy with conventional tuning and interchanged tuning. $vB_i$ ($vB_i'$) denotes the i-th virtual barrier gate voltage, while $f_{MW}$ corresponds to the microwave frequencies applied to the target qubit. $X$ ($X^2$) denotes a π/2 (π) rotation about the *x*-axis. $Q_C$ ($Q_C'$) and $Q_T$ ($Q_T'$) denote the control and target qubits, respectively, for each experiment.

**Figure 4.** Enhanced tunability of exchange coupling

**(a)** Decoupled Controlled-Z oscillations of the qubits by applying a varying pulse amplitude to the barrier gates. An $X$ or $X^2$ gates are performed on the $Q_C'$ (control qubit) or $Q_T'$ (target qubit) in the interchanged tuning strategy. A pulse with amplitude $\Delta vB_i'$ is applied to turn on the exchange coupling. For the echo sequence, the pulse is split into two equal halves of duration $t_{evol}/2$

**(b)** Exchange coupling as a function of the barrier pulse amplitude. The blue (orange) trace shows data from an interchanged (conventional) gate configuration. The extracted tunabilities are 16.6 dec/V, 11.4 dec/V, and 15.4 dec/V (7.25 dec/V, 3.87 dec/V, and 4.32 dec/V) for the interchanged (conventional) gate configuration, obtained by fitting to $J = Ae^{Bv}$. The error bars represents the 1σ uncertainties obtained from the fit.

# Supporting Information

## S1. Experimental setup

The quantum dot qubit device was mounted in a dilution refrigerator (Oxford Instruments Triton-500) with the mixing chamber plate kept at approximately 8 mK. All metal gates and ohmic contacts were connected to DC voltage sources (SIM928, Stanford Research Systems), with some additionally linked to either an arbitrary waveform generator (AWG, Quantum Machines OPX+ and SDT QCU) or a signal generator (Quantum Machines Octave) via on-board bias tees. The I/Q signals from the AWG were fed into the Octave, which functions as a local oscillator and up-converter, producing I/Q modulated microwave signals used to manipulate the spin qubits. Other baseband signals were employed to adjust the quantum dot energy levels or tunnel coupling abruptly. A lock-in amplifier (Zurich Instruments UHFLI) provided combined RF signals at the resonant frequencies of the two LC tank circuits, directed respectively to the two ohmic contacts. These RF signals, modulated by the conductance of the sensor dots, were reflected out and amplified by 45 dB through two series-connected cryogenic amplifiers (Cosmic Microwave Technology CITLF2), followed by an additional 25 dB of gain from a room-temperature amplifier (Lotus Communication Systems LNA100M2P0G). The amplified RF signals were then demodulated by a lock-in amplifier and digitized by the OPX+.

## S2. Initialization and measurement process of the two tuning strategies

With the conventional tuning strategy, the charge occupation was set to (3,1,1,1,3). To implement the Pauli-spin-blockade-based parity readout, we reduced the interdot tunneling coupling until triplet-state latching was observed in the charge stability diagram. The signal-to-noise ratios (SNRs) of the left and right sensor dots were calculated, provided that the measured signal data follow a double-Gaussian distribution. We assumed that the raw data were first binned to construct a histogram, and the count distribution $C(x)$ was fitted to extract parameters.

$$p(x) = w\frac{1}{\sqrt{2\pi}\sigma_1}e^{-\frac{(x-\mu_1)^2}{2\sigma_1^2}} + (1-w)\frac{1}{\sqrt{2\pi}\sigma_2}e^{-\frac{(x-\mu_2)^2}{2\sigma_2^2}}$$

$$C(x) = Np(x)\Delta x$$

Here, $\mu$, $\sigma$, $w$, $N$, and $\Delta x$ denote the mean, standard deviation, the population ratio of each state, the total number of samples, and the bin width, respectively. After that, the SNR is calculated with the following equation.

$$SNR = \frac{|\mu_1 - \mu_2|}{\sqrt{(\sigma_1^2 + \sigma_2^2)/2}}$$

The values are 19.6 and 9.25, with the integration time 1 μs and 2 μs, respectively, as shown in Figure S1a. We carried out the initialization procedure illustrated in Figure S1b to ensure that the two pairs of qubits had odd parity and that $Q_3$ was initialized in the spin-down state. During the “init1” process, if the measured parity was even, $Q_2$ or $Q_4$ in each pair was rotated by π. As shown in Figure S1c, we observed Rabi oscillations in all five qubits; however, the oscillations of $Q_3$ and $Q_5$ were not optimized. We attribute this to the imperfect initialization of $Q_3$ and the strong coupling with the nearby reservoir. Under this tuning strategy, the lever arm is extracted with the normalized signal across the charge transition line. This data is fitted to the Fermi-Dirac distribution to extract $\alpha / k_B T_e$, and the values are obtained for different temperatures of the mixing chamber. The lever arm is extracted to be 62.1 μeV/mV.

For the interchanged tuning strategy, the outermost qubits were placed farther from the neighboring sensor dot than before. Still, the SNRs of the sensors on the left and right (10.6 and 7.02, respectively) with an integration time of 2 μs remained sufficient to resolve the parity of the qubits (Figure S2a). Spin-qubit experiments used the initialization sequence shown in Figure S2b. The same "init1" process, where a conditional π-pulse was applied to $Q_2'$ or $Q_3'$, was used. An extra "init2" process employed a zCNOT gate on $Q_1'$ and $Q_4'$, based on controlled rotation, to reliably prepare the two-spin states. Then, we successfully initialized the $|0\rangle$ state, as shown by the results in Figure S2c, which features nearly a single resonance branch in the exchange spectroscopy measurement depending on the control qubit state. In each panel, the qubit symbols indicate the control qubits, and each panel displays the resonances of $Q_2'$, $Q_3'$, and $Q_4'$ from left to right.

From the charge stability diagrams measured in virtual gate voltages, we quantify the plunger voltage spacing required to load an extra electron into a given dot. The spacing is related to the additional energy through $E_{add} \approx \alpha \Delta V_P$, where $\alpha$ is a lever arm. Provided that the additional energy is comparable between the two tuning strategies, the ratio of lever arms can be estimated as $\frac{\alpha^{conv}}{\alpha^{inter}} \approx \frac{\Delta V_P^{inter}}{\Delta V_P^{conv}}$. We emphasize that this estimate is a proxy and may also include contributions from changes in $E_{add}$; therefore, we use it as a qualitative indicator of the reduced gate-to-dot coupling strength.

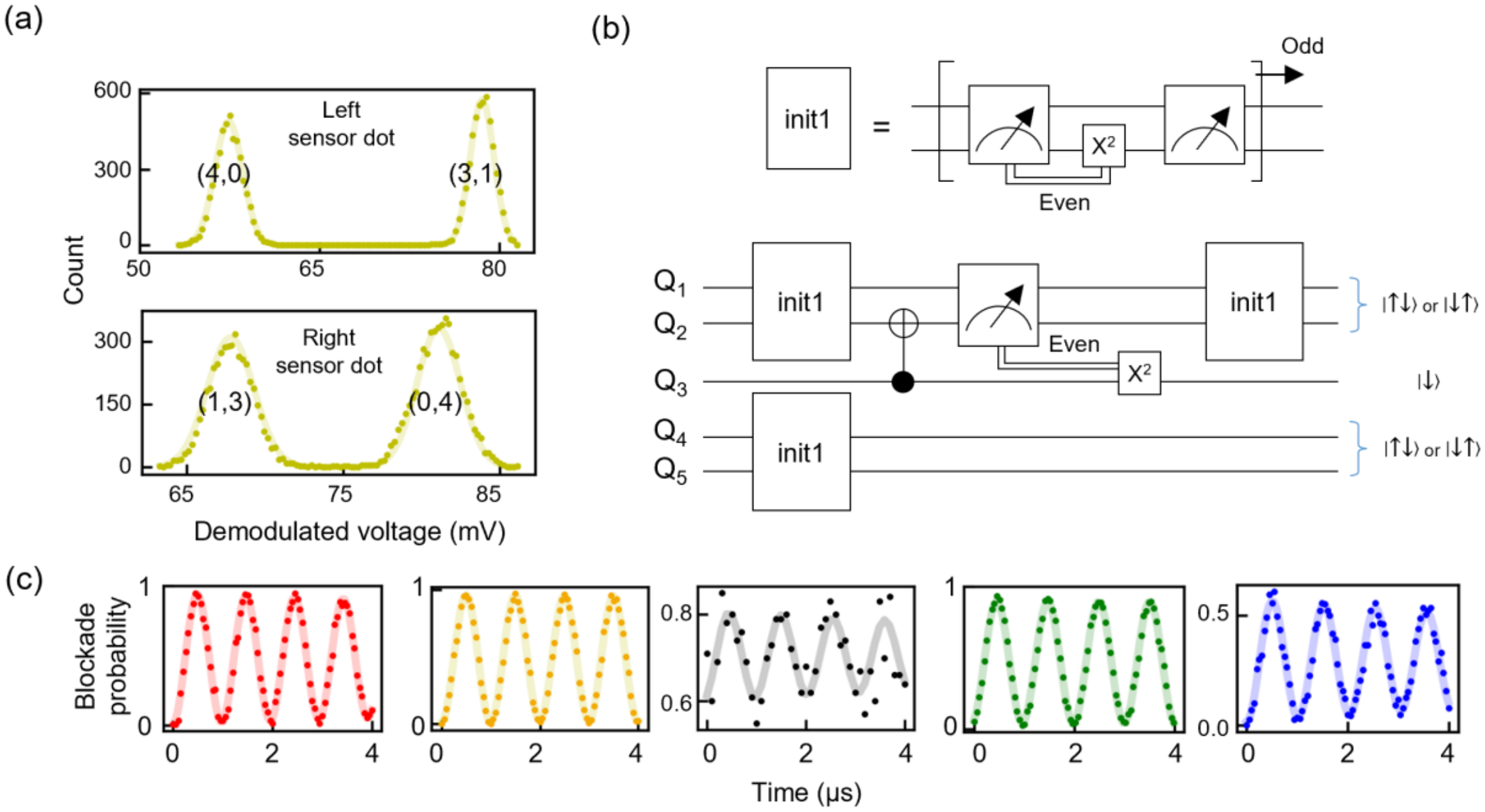

**Figure S1.** Initialization process and measurement using the conventional strategy

(a) Histogram of the digitized sensor dot signal with the conventional tuning strategy.

(b) Schematics of the initialization process for five-spin qubits.

(c) Rabi oscillations of the five spin qubits.

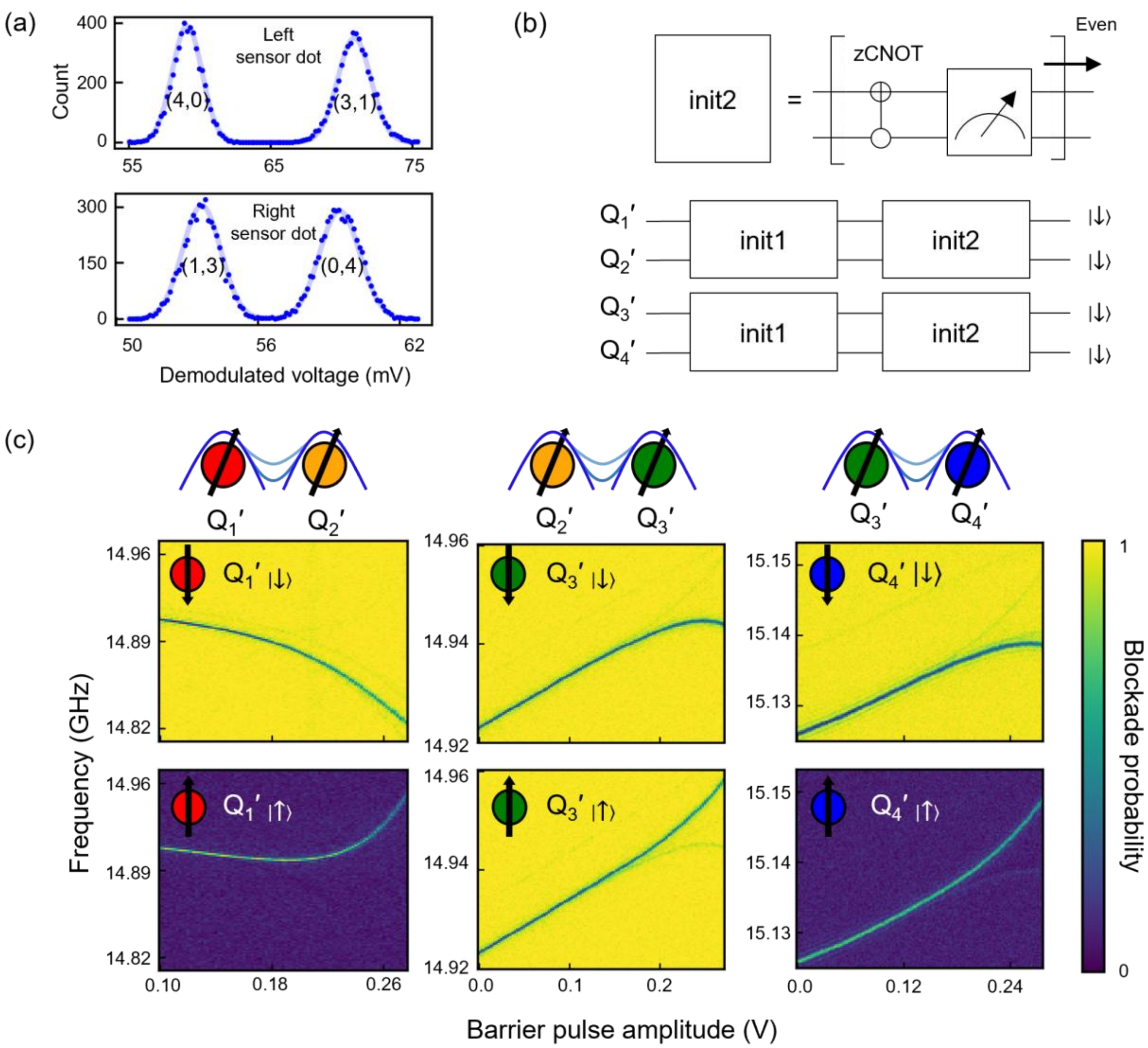

**Figure S2.** Initialization process and measurement with the interchanged tuning strategy

**(a)** Histogram of the digitized sensor dot signal.

**(b**) Schematics of the initialization process for four-spin qubits.

**(c)** Exchange spectroscopy measurements after the initialization process.

### S3. Simulation of electric potential and tunnel coupling

The electrostatic confinement potential and tunnel coupling were simulated prior to experimental validation of the interchanged tuning strategy. The layer stack and individual layer thicknesses of the $^{28}$Si/SiGe wafer adhered to the specifications reported in a previous study.[1] The morphology of the gate oxide layer ($Al_2O_3$ 20 nm), metal gate layers (Ti 5 nm, Pd 30 nm), and intervening oxide layers ($Al_2O_3$ 8 nm) were reflected in the simulation via observing a cross-sectional transmission electron microscope image of the device prepared in the same fabrication batch as the device used in the experiment. We employed COMSOL MULTIPHYSICS to generate an electric potential at the location of the two-dimensional electron gas (2DEG), which was estimated to reside at the interface between the $^{28}$Si well and the SiGe barrier layer. After specifying the boundary conditions and gate voltages, we solved Poisson's equation self-consistently with the 2DEG charge density within the Thomas-Fermi approximation [2]

$$\rho_{2DEG}(V) = -e\frac{g_v m_t^*}{\pi\hbar^2}(E_F + eV)\Theta(E_F + eV) \qquad (1)$$

Here, $g_v$ denotes the valley degeneracy, $m_t^*$ is the transverse electron effective mass, and $\Theta(x)$ is the Heaviside step function. In this simulation, we set $g_v = 2$ and $E_F = 0$. By varying the voltage levels applied to the barrier gate (*vB3* and *vB3'*), we obtained the two-dimensional electrostatic potential maps and subsequently solved the single-particle Hamiltonian in a maximally localized basis to extract the off-diagonal tunnel-coupling term.

## S4. Virtual gate configuration

Each metal gate affects not only the quantum dot beneath it but also neighboring dots, making independent control challenging. To control the energy of a specific quantum dot, virtual gates are commonly employed to suppress crosstalk.[3] We investigated the cross-capacitance matrix for both tuning strategies to achieve improved control of the quantum dots. Metal gates located near a quantum dot also contribute to shifts in its energy, which appear as non-zero off-diagonal elements in Figure S3. In this work, the compensation is applied to the electrochemical potential of quantum dots, but not to the tunnel couplings.

During the initial quantum dot tuning (Figure 2), physical gates were used. In subsequent experiments (Figure 3), we employed the virtual gates. As the newly assigned plunger gates are spaced farther from 2DEG than in the conventional tuning strategy, higher voltages are required for detuning ramps; however, the cross-capacitance between the plungers remains manageable. Although the influence of the newly assigned barrier gates has increased, reliable two-qubit interaction is still achieved even at the maximum barrier pulse amplitude, as shown in Figure 3, without observing any unwanted transition lines that could arise from insufficient cross-capacitance compensation.

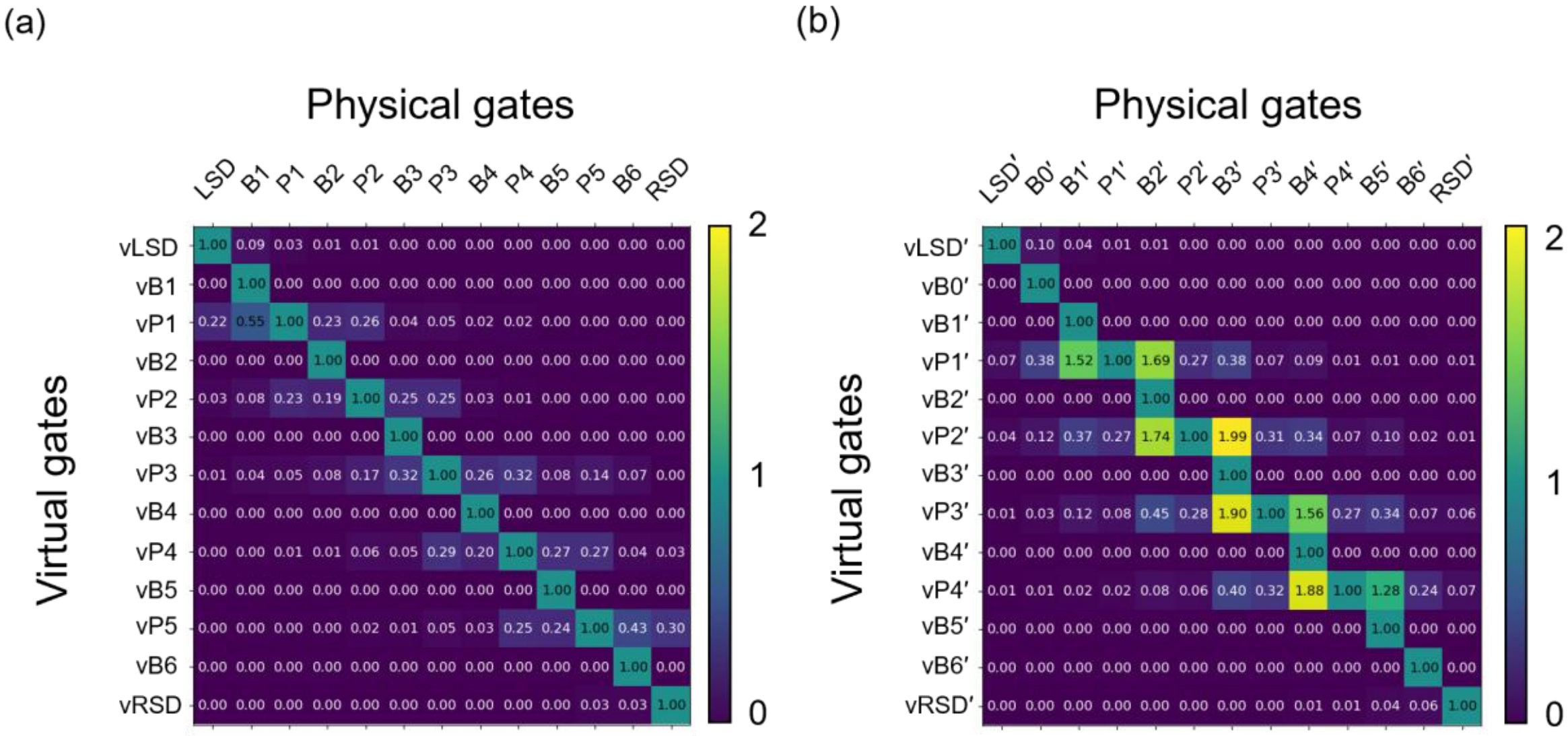

**Figure S3. Visualization of cross-capacitance matrices**

The matrix for **(a)** Conventional tuning strategy and **(b**) Interchanged tuning strategy

## S5. Exchange oscillation quality between two tuning strategies

We further investigate the quality of exchange oscillations for the two tuning strategies. Figure S4a shows the decoupled CZ oscillation for the $Q_1$-$Q_2$ pair using a conventional tuning strategy, which is obtained by varying the barrier pulse amplitude and evolution time. Figure S4b and S4c show the trend of decay time and exchange coupling as a function of barrier pulse amplitude. As the coupling strength increases exponentially, the decay time correspondingly decreases. The oscillation quality can be quantified by the product of the decay time and the coupling strength. We extract the quality factor and plot it as a function of exchange coupling for both tuning strategies. (Figure S4d) When the exchange coupling is small, the decay time is long;

however, the oscillation is too slow, resulting in a low quality factor for both strategies. As the exchange coupling becomes stronger, the quality factor increases significantly due to the exponential growth of the coupling strength. The quality factor of the conventional tuning strategy fluctuates around ~15 beyond 5 MHz of exchange coupling. For exchange coupling beyond 30 MHz, the decay time becomes too short to clearly resolve the oscillations. Notably, the quality factor of the interchanged tuning strategy is comparable to, or slightly higher than, that of the conventional tuning strategy for some exchange coupling strengths. The result demonstrates that the interchanged tuning strategy does not degrade the two-qubit operation based on exchange coupling.

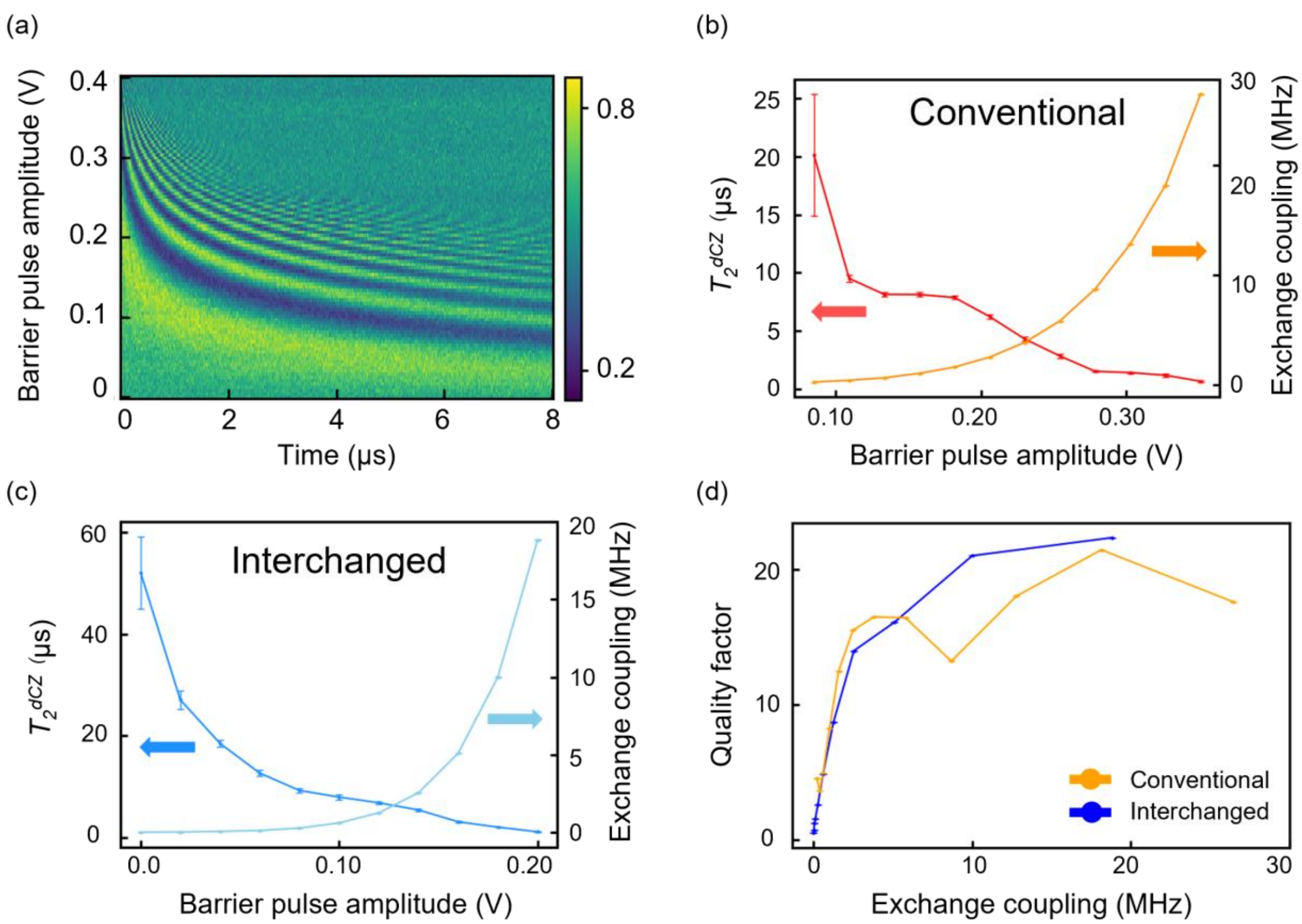

**Figure S4. Comparison of quality factors between two strategies**

**(a)** Decoupled CZ oscillation for $Q_1$-$Q_2$ pair with the conventional tuning strategy

**(b-c)** Decay time and exchange coupling as a function of the barrier pulse amplitude for conventional and interchanged tuning strategies, respectively. The error bars indicate the 1σ uncertainties obtained from the fits.

**(d)** Quality factor versus exchange coupling. The error bars correspond to the 1σ uncertainties obtained from the fits.

## S6. Effect of metal gate geometry on lever arm and exchange coupling tunability

To further investigate the effects of gate geometry on lever-arm and exchange-coupling tunability, we performed numerical simulations. The electrostatic potential was calculated using COMSOL Multiphysics, followed by solving the single-particle Hamiltonian (see Section S3). The lever arm was extracted from the shift in the ground-state orbital energy as a function of the applied gate voltage. The exchange coupling was obtained using a full configuration interaction (FCI) approach, in which the two-electron states were constructed, and the exchange coupling strength was defined as the energy difference between the ground singlet and triplet states.[4] The exchange coupling tunability was then evaluated by tracking the variation in the coupling strength as a function of the gate voltage.

Figure S5a shows a schematic cross-section of the device with rectangular metal gates having a width of 30 nm, a thickness of 35 nm, and a pitch of 90 nm. Figure S5b shows another gate geometry, a trapezoidal shape with the same pitch and thickness as in Figure S5a, but with top widths of 20 (40) nm and bottom widths of 40 (20) nm for the second (third) layer gates. In our device, the fabricated gate shape is more trapezoidal than rectangular (Figure 1b-c) due to the lift-off process using electron beam lithography and electron beam evaporation. The vertical separation between the two gate layers was fixed at 8 nm, corresponding to the thickness of the intervening oxide layer used in the device.

Figure S5c shows the simulated lever arm and exchange coupling tunability. The marker color indicates the gate layer, the marker shape indicates the gate geometry, and the marker filling reflects the distance (*D*) between the silicon oxide layer and the bottom of the second gate layer. We investigated two layers with rectangular and trapezoidal shapes at varying *D*. As expected, the exchange coupling tunability shows a linear dependence on lever arm, and the lever arm increases as *D* decreases. In addition, we find that the difference in lever arm between the second and third layers becomes more pronounced for trapezoidal gates. Although the gates were designed to have the same pitch and width, the fabrication process modifies the gate shape, which in turn affects lever arm and exchange coupling tunability. These results indicate that the interchange tuning strategy can be effectively employed with trapezoidal metal gates, thereby enhancing capacitive coupling between the second-layer gate (green) and the quantum dot. In conclusion, the design of lever arm and exchange coupling tunability should take into account not only the gate width and pitch, but also the effective bottom width, which can arise from a thick oxide layer or from a small pitch-to-width ratio.

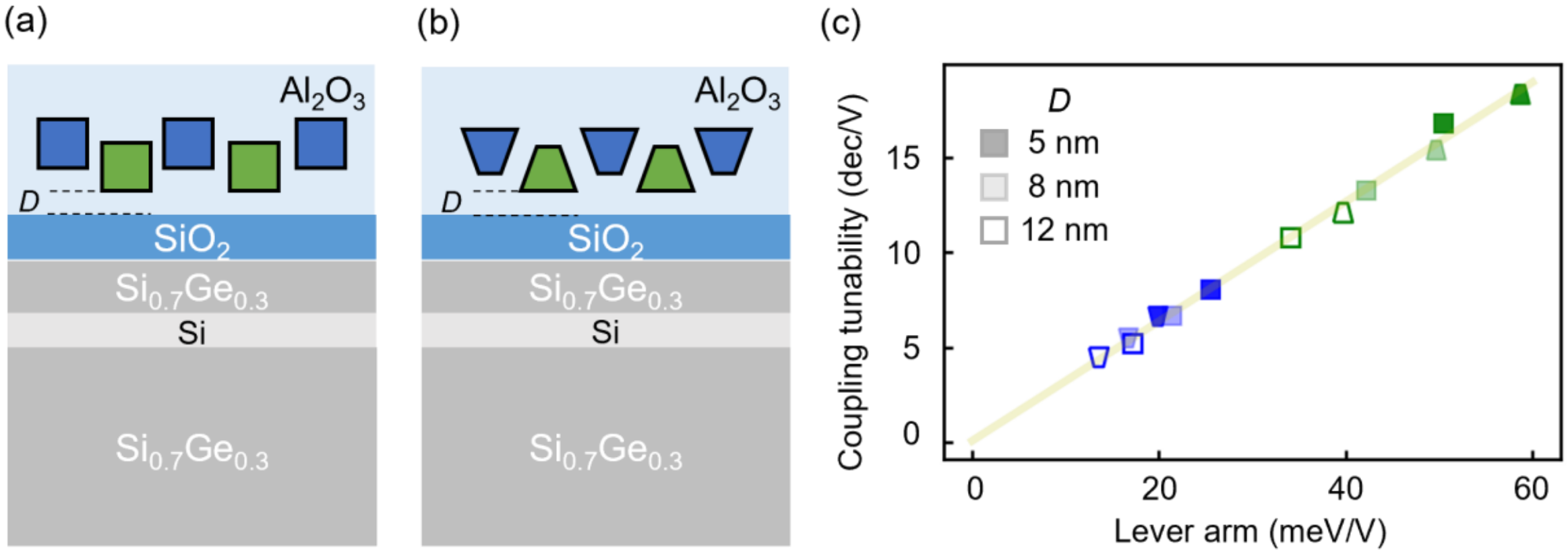

**Figure S5.** Simulation results as a function of the metal gate geometry.

**(a-b)** Schematics of the device cross-section with square and trapezoidal metal gates, respectively.

**(c)** Simulated lever arm and exchange coupling tunability.

## Supplementary references